\documentclass[aip,reprint,nofootinbib]{revtex4-1}
\usepackage{graphicx}
\usepackage{bm}

\begin{document}
\title{A Geometrical Method of Decoupling}
\date{\today}
\author{C. Baumgarten}
\affiliation{Paul Scherrer Institute, Switzerland}
\email{christian.baumgarten@psi.ch}

\def\begeq{\begin{equation}}
\def\endeq{\end{equation}}
\def\begary{\begeq\begin{array}}
\def\endary{\end{array}\endeq}
\def\bmtx{\left(\begin{array}}
\def\emtx{\end{array}\right)}
\def\eps{\varepsilon}
\def\d{\partial}
\def\y{\gamma}
\def\e{\eta}
\def\w{\omega}
\def\W{\Omega}
\def\s{\sigma}
\def\ket#1{\left|\,#1\,\right>}
\def\bra#1{\left<\,#1\,\right|}
\def\bracket#1#2{\left<\,#1\,\vert\,#2\,\right>}
\def\erw#1{\left<\,#1\,\right>}

\def\Exp#1{\exp\left(#1\right)}
\def\Log#1{\ln\left(#1\right)}
\def\Sinh#1{\sinh\left(#1\right)}
\def\Sin#1{\sin\left(#1\right)}
\def\Tanh#1{\tanh\left(#1\right)}
\def\Tan#1{\tan\left(#1\right)}
\def\Cos#1{\cos\left(#1\right)}
\def\Cosh#1{\cosh\left(#1\right)}

\begin{abstract}
The computation of tunes and matched beam distributions are essential steps
in the analysis of circular accelerators. If certain symmetries -- like midplane 
symmetrie -- are present, then it is possible to treat the betatron motion in 
the horizontal, the vertical plane and (under certain circumstances) the longitudinal 
motion separately using the well-known Courant-Snyder theory, or to apply 
transformations that have been described previously as for instance the
method of Teng and Edwards~\cite{Teng,EdwardsTeng}. In a preceeding paper
it has been shown that this method requires a modification for the treatment 
of isochronous cyclotrons with non-negligible space charge forces~\cite{cyc_paper}.
Unfortunately the modification was numerically not as stable as desired and
it was still unclear, if the extension would work for all conceivable cases. 
Hence a systematic derivation of a more general treatment seemed advisable.

In a second paper the author suggested the use of real Dirac matrices as basic
tools for coupled linear optics and gave a straightforward recipe to decouple 
positive definite Hamiltonians with imaginary eigenvalues~\cite{rdm_paper}. 
In this article this method is generalized and simplified in order to formulate 
a straightforward method to decouple Hamiltonian matrices with eigenvalues on 
the real and the imaginary axis. The decoupling of symplectic matrices which are 
exponentials of such Hamiltonian matrices can be deduced from this in a few steps. 
It is shown that this algebraic decoupling is closely related to a  geometric 
``decoupling'' by the orthogonalization of the vectors $\vec E$, $\vec B$ and $\vec P$, 
that were introduced with the so-called ``electromechanical equivalence''~\cite{rdm_paper}.

A mathematical analysis of the problem can be traced down to the task of
finding a structure-preserving block-diagonalization of symplectic or Hamiltonian
matrices. Structure preservation means in this context that
the (sequence of) transformations must be symplectic and hence canonical.

When used iteratively, the decoupling algorithm can also be applied to 
n-dimensional systems and requires ${\cal O}(n^2)$ iterations to converge to 
a given precision.
\end{abstract}

%Hamiltonian mechanics, 45.20.Jj
%Oscillators coupled, 05.45.Xt
%Beam optics  41.85.-p
%Lorentz transformation 03.30.+p
\maketitle

%%%%%%%%%%%%%%%%%%%%%%%%%%%%%%%%%%%%%%%%%%%%%%%%%%%%%%%%%%%%%%%%%%%%%%%%%%%%%%%%
\section{Introduction}
%%%%%%%%%%%%%%%%%%%%%%%%%%%%%%%%%%%%%%%%%%%%%%%%%%%%%%%%%%%%%%%%%%%%%%%%%%%%%%%%

The significance of the symplectic groups in Hamiltonian dynamics has been emphasized 
for instance by A. Dragt~\cite{Dragt}, and it has long been known~\cite{DGL} that the 
Dirac matrices are generators of the symplectic group $Sp(4,R)$.
In Ref.~\cite{rdm_paper} the author presented 
a toolbox for the treatment of two coupled harmonic oscillators that is based on 
the use of the real Dirac matrices (RDMs) as generators of the symplectic group 
$Sp(4,R)$ and a systematic survey of symplectic transformations in two dimensions. 
This toolbox enabled the developement of a straightforward recipe for the decoupling of 
positive definite two-dimensional harmonic oscillators. Here we present an 
improvement of the method that is based on geometric arguments, i.e. on the 
orthogonalization of 3-dimensional vectors associated via the electromechanical 
equivalence (EMEQ) to certain linear combinations of matrix elements.

There is a long history of publications covering the diagonalization (and related) 
problems in linear algebra as well as in linear coupled optics, linear Hamiltonian 
dynamics and control theory. A (non-exhaustive) list is given in the 
bibliography (see Refs.~\cite{LMC,PvL,vanLoan,BMX,BMX2,Coleman,Dieci,SR,YQXX,
BKM,Luo,CM,MK,MSW,ABK,BFS,CS}, but also Ref.~\cite{cyc_paper,rdm_paper} and 
references therein). However, none of the previous works (known to the author) 
takes full advantage of the group structure of the generators of $Sp(4)$. 
The conceptually closest approach uses ``quaternions'', the representations of which 
seems to be identical to the RDMs~\cite{FMM}, but seems to be limited to 
orthogonal symplectic transformations.
The decoupling method of Teng and Edwards has been the starting point for this work, 
as it turned out to fail in some special cases (see Ref.~\cite{cyc_paper} and App.~\ref{sec_ET}). 

The method that we present here, is based on a survey of all symplectic similarity
transformations. We do not make specific assumptions about the Hamiltonian other than 
that it is a symmetric quadratic form and we present a geometric interpretation via the EMEQ, 
which provides a physical notation of otherwise complicated and non-descriptive
algebraic expressions~\footnote{Compare for instance Ref.~\cite{CMC}.}. 
Furthermore we believe that the use of the EMEQ is an interesting example of how 
elements of classical physics, quantum mechanics, special relativity, electrodynamics, 
group theory, geometric algebra, statistics~\cite{stat_paper} and last but not least 
symplectic theory fit together and allow to use a common formalism.

The simplest classical linear dynamical system with interaction (coupling) has 
two degrees of freedom and hence a 4-dimensional phase space. It can be considered 
as ``fundamental'' and a detailed analysis of its properties will
likely be instructive also for $n>2$. Indeed it turns out, that the decoupling
technique of a two-dimensional system can be iteratively applied to systems
with more than two degrees of freedom. A Jacobi-like iteration with
pivot-search and is sketched in Sec.~\ref{sec_2nx2n}.

From the viewpoint of coupled linear optics the problem is solved if a 
symplectic transformation is derived that transforms (constant) Hamiltonian
matrices to $2\times 2$-block-diagonal form (see below). It has been shown in
Ref.~\cite{rdm_paper}, that the same transformation method can be applied to
symplectic matrices as well. The arguments will be briefly reported below.
When applied to symplectic matrices, the method is equivalent to the computation 
of the matrix logarithm. A solution for the couterpart, i.e. the computation
of the matrix exponential with emphasis on the use of Dirac matrices, 
has been presented by Barut, Zeni and Laufer in 1994~\cite{BZL}.

If $2\times 2$-block-diagonal form has been achieved, the remaining task is 
completely analogous to the application of the Courant-Snyder theory for one 
degree of freedom. 
Nevertheless some arguments require awareness of the eigenvalues and their
relation to the properties of the Dirac matrices so that a reference to a 
complete diagonalization seemed appropriate. 

\section{Coupled Linear Optics}
\label{sec_CLO}

The Hamiltonian of a $n$-dimensional harmonic oscillator with arbitrary 
coupling terms can be written in the form
\begeq
H={1\over 2}\,\psi^T\,{\bf A}\,\psi\,,
\label{eq_Hamiltonian}
\endeq
where ${\bf A}$ is a symmetric matrix and $\psi$ is a state-vector
or ``spinor'' of the form $\psi=(q_1,p_1,q_2,p_2,\dots,q_n,p_n)^T$. Even though the matrix 
${\bf A}$ is time-dependent in the general case, it is well-known practice
to use the Floquet-transformation to reduce it to a constant matrix
for the treatment of periodic systems (see app.~\ref{app_floquet} and 
for instance Ref.~\cite{Talman,MHO}).
The symplectic unit matrix (usually labeled ${\bf J}$ or ${\bf S}$) is a skew-symmetric
matrix that squares to the negative unit matrix. For $n=2$ it is identified 
with the real Dirac matrix $\y_0$.
As described in Ref.~\cite{rdm_paper}, it is possible to freely choose the 
order of the variables in the state vector. However, the order of the variables
fixes the form of the symplectic unit matrix $\y_0$~\footnote{
For $n=2$, this also follows from the fundamental theorem
of the Dirac matrices (see for instance Ref.~\cite{AJM,DGL}).}.
We prefer the use of a ordering system in which the phase space coordinates 
$(q_i,p_i)$ are grouped as pairs of canonical conjugate variables, so that
$\y_0$ has the form
\begeq
\y_0=\bmtx{ccccc}
 0 & 1 &      &0 &0 \\
-1 & 0 &\dots &0 &0 \\
   & \vdots&&\vdots&\\
0  & 0 &\dots & 0 & 1 \\
0  & 0 &      &-1 & 0 \\
\emtx\,,
\label{eq_gamma0}
\endeq

Using the over-dot to indicate the derivative with respect to time (or path length), 
the equations of motion (EQOM) have the familiar form 
\begary{rclp{5mm}rcl}
\dot q_i&=&{\d H\over \d p_i}&&\dot p_i&=&-{\d H\over \d q_i}\,,
\label{eq_eqom_classical}
\endary
or in vector notation:
\begeq
\dot\psi=\y_0\,\nabla_\psi\,H={\bf F}\,\psi\\
\label{eq_eqom_general}
\endeq
where the {\it force matrix} ${\bf F}$ is given as  
\begeq
{\bf F}=\y_0\,{\bf A}\,.
\label{eq_Fdef}
\endeq
From the definition of ${\bf F}$ one quickly finds that~\cite{Talman,MHO}
\begeq
{\bf F}^T=\y_0\,\,{\bf F}\,\y_0\,,
\label{eq_symplex}
\endeq
where the superscript ``T'' denotes the transposed matrix.
Matrices that obey Eqn.~(\ref{eq_symplex}) are usually called
``infinitesimally symplectic'' or ``Hamiltonian''~\cite{Talman}. 
Both terms are - in the opinion of the author - misleading: The former 
because ${\bf F}$ is neither symplectic nor is it infinitesimal, the latter 
since ${\bf F}$ does not appear in the Hamiltonian while the symmetric matrix
${\bf A}$ does. In addition ${\bf A}$ and not ${\bf F}$ is in the view of the
author the classical counterpart of the Hamiltonian operator 
(see App. A in Ref.~\cite{rdm_paper}).
Furthermore, Eqn.~(\ref{eq_symplex}) is a purely formal property and not necessarily
connected to a Hamiltonian. Therefore the author decided to use the term ``symplex'' 
(plural ``symplices'') when referring to its formal definition (i.e. Eqn.~(\ref{eq_symplex})) 
and its relation to the symplectic transfer matrix and to call it ``force matrix'' 
when referring to its physical content - especially with respect to the 
EMEQ (see Ref.~\cite{rdm_paper} and below).
Accordingly we speak of an anti-symplex or {\it cosymplex} (i.e. ``skew-Hamiltonian'' matrix),
if a matrix ${\bf C}$ fulfills the equation 
\begeq
{\bf C}^T=-\y_0\,\,{\bf C}\,\y_0\,.
\label{eq_cosymplex}
\endeq
If we write ${\bf S}$ (${\bf C}$) for (co-)symplices, respectively, optionally with a subscript, 
then it is easy to prove that
\begeq
\left.\begin{array}{c}
{\bf S}_1\,{\bf S}_2-{\bf S}_2\,{\bf S}_1\\
{\bf C}_1\,{\bf C}_2-{\bf C}_2\,{\bf C}_1\\
{\bf C}\,{\bf S}+{\bf S}\,{\bf C}\,,
\end{array}\right\} \Rightarrow\mathrm{symplex}\,,
\endeq
and 
\begeq
\left.\begin{array}{c}
{\bf S}_1\,{\bf S}_2+{\bf S}_2\,{\bf S}_1\\
{\bf C}_1\,{\bf C}_2+{\bf C}_2\,{\bf C}_1\\
{\bf C}\,{\bf S}-{\bf S}\,{\bf C}\\
\end{array}\right\}\Rightarrow\mathrm{cosymplex}\,.
\endeq

\subsection{Dirac Matrices}

In the following we focus on two degrees of freedom ($n=2$), i.e. to a four-dimensional
phase space and the use of the real Dirac matrices to describe its dynamics and transformation
properties. 
Often the term ``Dirac matrices'' is used more restrictively and designates only four matrices,
namely $\y_k\,,\,k\in\,[0\dots 3]$. Here we consider the four basic Dirac matrices as the
four basic elements of a Clifford algebra $Cl(3,1)$ with $16$ elements derived from the basic 
matrices (see app.~\ref{sec_app1}). For further details see for instance Ref.~\cite{Okubo,Scharnhorst,Hestenes}.

Any real $4\times 4$-matrix ${\bf M}$ can be written as a linear combination of the RDMs
\begeq
{\bf M}=\sum\limits_{k=0}^{15}\,m_k\,\y_k\,.
\endeq
The RDM-coefficients $m_k$ are given by~\footnote{Eqn.~(\ref{eq_rdmcoeffs}) is based
on the fact that all RDMs except the unit matrix have zero trace.}
\begeq
m_k=\mathrm{Tr}(\y_k^2)\,\mathrm{Tr}\left({{\bf M}\,\y_k+\y_k\,{\bf M}\over 32}\right)\,,
\label{eq_rdmcoeffs}
\endeq
where $\mathrm{Tr}({\bf X})$ is the trace of the matrix ${\bf X}$. 
Only the first ten RDMs are symplices and since symplices obey the superposition 
principle~\cite{rdm_paper,MHO,Dragt}, any force matrix (symplex) can be written as
\begeq
{\bf F}=\sum\limits_{k=0}^{9}\,f_k\,\y_k\,.
\label{eq_force_coeffs}
\endeq
The solution of Eqn.~(\ref{eq_eqom_general}) is known to be
\begeq
\psi(s)=\exp{({\bf F}\,s)}\,\psi(0)\,,
\label{eq_transfer}
\endeq
where the matrix 
\begeq
{\bf M}=\exp{({\bf F}\,s)}
\label{eq_expo}
\endeq
is called {\it transfer matrix}, which can be shown to fulfill the symplectic
condition, if ${\bf F}$ is a symplex~\cite{rdm_paper,Dragt,MHO}:
\begeq
{\bf M}\,\y_0\,{\bf M}^T=\y_0\,.
\label{eq_symplectic}
\endeq
Vice versa it is known that symplectic matrices
can be written in the form of Eqn.~(\ref{eq_expo})~\cite{Talman,MHO}. 

Transfer matrices can be split into two parts, one ($M_s$) being
a symplex, the other ($M_c$) being a cosymplex~\cite{rdm_paper,Parzen}:
\begary{rcl}
{\bf M}_c&=&({\bf M}-\y_0\,{\bf M}^T\,\y_0)/2\\
{\bf M}_s&=&({\bf M}+\y_0\,{\bf M}^T\,\y_0)/2\,,
\label{eq_split}
\endary
which is in case of a symplectic matrix ${\bf M}$ identical to
\begary{rcl}
{\bf M}_c&=&({\bf M}+{\bf M}^{-1})/2\\
{\bf M}_s&=&({\bf M}-{\bf M}^{-1})/2\,.
\endary
It has been shown in Ref.~\cite{rdm_paper}, that the decoupling of
the symplex-part ${\bf M}_s$ of a symplectic matrix ${\bf M}$ 
automatically decouples the corresponding cosymplex ${\bf M}_c$.
Hence it is sufficient to derive a method to decouple symplices of 
the above mentioned type.
In cases where only the one-turn-transfer matrix is available, 
Eqn.~(\ref{eq_split}) is used beforehand to extract the symplex-part 
of the transfer matrix. The decoupling algorithm can then be applied
to this matrix (see also the detailed discussion in Ref.~\cite{rdm_paper}).

\section{Block-Diagonalization and Eigenvalues}

The force matrix ${\bf F}$ is by definition a product of a symmetric
matrix ${\bf A}$ and of a skew-symmetric matrix $\y_0$. Hence it has
zero trace and the sum of all eigenvalues is zero.
We restrict ourselves to systems with real-valued force matrices
and therefore real-valued transfer matrices. The eigenvalues of
real-valued $2\times 2$-symplices are either both real
or both purely imaginary (since they are the square root of a real expression). 
Block-diagonalization (in the case of the variable ordering as described 
above) means to find a symplectic similarity transformation ${\bf R}$ 
such that the matrix ${\bf\tilde F}={\bf R}\,{\bf F}\,{\bf R}^{-1}$ has the form
\begeq
{\bf\tilde F}=\bmtx{cc}
{\bf\tilde F}_1&0\\
0&{\bf\tilde F}_2\emtx\,,
\endeq
where ${\bf\tilde F}_k$ are real $2\times 2$-matrices. Since similarity 
transformations preserve the eigenvalues, a symplex is 
block-diagonalizable in the form that we are going to describe, 
if the (pairs of) eigenvalues are either real or imaginary.
In case of imaginary eigenvalues, the corresponding degree of freedom
(i.e. pair $(q_i,p_i)$) is stable (or focused), while a pair of real eigenvalues
belongs to an unstable (non-focused) degree of freedom.
The corresponding betatron motion is unstable in the sense, 
that no sufficient focusing is present.

However -- in the general coupled case without further assumptions --
${\bf F}$ is a general $4\times 4$-symplex (or larger). Using the RDMs 
it is relatively easy to construct matrices with complex eigenvalues. 
An example is 
\begeq
{\bf F}=E_x\,\y_4+B_x\,\y_7\,,
\endeq
which has the complex eigenvalues $\pm i\,(B_x\pm i\,E_x)$.
Since the eigenvalues are complex, also the $2\times 2$-blocks
are complex.
They can be block-diagonalized, but the generalization to the 
$2\,n\times 2\,n$-case requires a general treatment of the complex case, 
which goes beyond the scope of this paper.

As in Ref.~\cite{rdm_paper} the author speaks of {\it regular} or {\it
  massive} systems, if the Hamiltonian is positive definite and of 
{\it irregular} or {\it magnetic} systems in case of indefinite Hamiltonian, 
respectively. Both types may be stable or unstable and this distinction
should not be confused with the question of stability. A detailed 
discussion of stability would go beyond the scope of this paper and we
refer the reader for instance to Ref.~\cite{MHO} or Ref.~\cite{FMM}
and references therein.

\subsection{The $\bf S$-matrix}

The matrix of second moments $\sigma$ of a charged particle distribution
\begeq
\sigma=\langle\psi\,\psi^T\rangle\,,
\label{eq_sigmadef}
\endeq
has the time derivative
\begeq
\dot\sigma={\bf F}\,\sigma+\sigma\,{\bf F}^T\,.
\endeq
Multiplication from the left with $\y_0$ and the use of Eqn.~(\ref{eq_symplex})
leads to 
\begeq
{\bf\dot S}={\bf F}\,{\bf S}-{\bf S}\,{\bf F}\,,
\endeq
where the matrix ${\bf S}$ is defined by
\begeq
{\bf S}=\sigma\,\y_0\,.
\label{eq_Smatrix}
\endeq
If Eqn.~(\ref{eq_Smatrix}) is compared to Eqn.~(\ref{eq_Fdef}), then
it is obvious that ${\bf S}$ is also a symplex as it is also the product
of a symmetric and a skew-symmetric matrix and obeys Eq.~\ref{eq_symplex}.
From Eqns.~(\ref{eq_transfer}), (\ref{eq_expo}) and (\ref{eq_sigmadef}) it follows that
\begeq
\sigma(s)={\bf M}(s)\,\sigma(0)\,{\bf M}^T(s)\,.
\endeq
The second moments of a matched distribution are unchanged after one turn 
(or sector) of period $L$ so that $\sigma(L)=\sigma(0)$ so that one obtains
in a few steps~\footnote{See common textbooks on linear Hamiltonian
dynamics or Ref.~\cite{rdm_paper}.}:
\begeq
{\bf M}\,{\bf S}-{\bf S}\,{\bf M}=0\,.
\endeq

\subsection{The Eigensystems and Matching}

Hence one finds that the matrices ${\bf M}$, ${\bf F}$ and ${\bf S}$ have
the same eigenvectors - but in general different eigenvalues~\cite{HMMG,Wolski}:
\begary{rclp{5mm}rcl}
{\bf F}&=&{\bf E}\,\lambda\,{\bf E}^{-1}&&{\bf M}&=&{\bf E}\,\Lambda\,{\bf E}^{-1}\\
{\bf S}&=&{\bf E}\,{\bf D}\,{\bf E}^{-1}&&\\
\label{eq_eigen}
\endary
where~\cite{Wolski}
\begary{rcl}
\lambda&=&\mathrm{Diag}(i\,\w_1,-i\,\w_1,i\,\w_2,-i\,\w_2)\\ 
\Lambda&=&\mathrm{Diag}(e^{i\,\w_1},e^{-i\,\w_1},e^{i\,\w_2},e^{-i\,\w_2})\\ 
{\bf D}&=&\mathrm{Diag}(-i\,\eps_1,i\,\eps_1,-i\,\eps_2,i\,\eps_2)\,.
\label{eq_evalues}
\endary
$\w_i$ are the oscillation frequencies and $\eps_i$ the emittances.
If $\bf E$ is known, the second moments of the matched distribution can 
be computed by replacing the eigenfrequencies by the emittances.
If a sympletic transformation ${\bf R}$ is known, that brings ${\bf F}$
(and hence ${\bf S}$ and ${\bf M}$) to block-diagonal form, then one
can simply use the usual Courant Snyder theory for one-dimensional 
systems~\cite{Hinterberger}.
In this case an explicit computation of the eigenvectors is not required.

\section{The Electromechanical Equivalence}

It was shown in Ref.~\cite{rdm_paper}, that the ten coefficients of the 
force matrix ${\bf F}$ or the ${\bf S}$-matrix can be identified with 
energy ${\cal E}$ and momentum $\vec P$ {\it of} a particle and with 
electric and magnetic field ($\vec E$ and $\vec B$, respectively) 
{\it seen by} a charged particle in external fields. The meaning of
this identification is, that the corresponding coefficients of ${\bf F}$ or ${\bf S}$ 
transform under symplectic transformations in the exact same way as 
the fields and the momentum transform under the corresponding boosts and 
rotations.

It was also shown that the {\it envelope equations} of coupled linear optics 
are isomorphic to the Lorentz force equation. The Lorentz group was found to 
be a subset of the two-dimensional symplectic group. The so defined ``fields''
($\vec E$ and $\vec B$) of the EMEQ should not be confused with the {\it real} 
fields of the beamline elements or accelerator components.

This isomorphism has been named {\it electromechanical equivalence} (EMEQ). 
The ten possible symplectic transformations are identified with spatial
and phase-rotations, Lorentz boosts and so-called ``phase boosts''. The 
transformation properties are analogous to those in Minkowski space-time.

This structural analogy is the basic idea behind the electromechanical equivalence (EMEQ).
Naturally, $\y_0$ is associated with the time-like components of 4-vectors (i.e. energy),
the spatial matrices $\vec\y=(\y_1,\y_2,\y_3)^T$ are associated with the momentum,
the matrices $\y_0\,\vec\y$ with the electric field and $\y_{14}\,\y_0\,\vec\y$ with
the magnetic field. The pseudoscalar has been named $\y_{14}=\y_0\,\y_1\,\y_2\,\y_3$ 
(instead of $\y_5$, as convention in QED). The remaining six matrices
are $\y_{10}$, which is the time-component of the pseudo-vector, 
$(\y_{11},\y_{12},\y_{13})^T=\y_{14}\,\vec\y$ are the spatial components of the
pseudo-vector and $\y_{15}={\bf 1}$ is the unit matrix. A complete list is given in
App.~\ref{sec_app1}, further details in Ref.~\cite{rdm_paper} and in textbooks on
quantum electrodynamics.

The EMEQ is given by the following nomenclature:
\begary{rcl}
{\cal E}&\equiv&f_0\\
\vec P&\equiv&(f_1,f_2,f_3)^T\\
\vec E&\equiv&(f_4,f_5,f_6)^T\\
\vec B&\equiv&(f_7,f_8,f_9)^T\,,
\endary
with the $f_k$ given by Eqn.~(\ref{eq_force_coeffs}).
Using the EMEQ, the eigenvalues of ${\bf F}$ (Eqn.~\ref{eq_eigen} and Eqn.~\ref{eq_evalues}) 
can be expressed by:
\begary{rcl}
K_1&=&{\cal E}^2+\vec B^2-\vec E^2-\vec P^2\\
K_2&=&-2\,{\cal E}\,\vec P\cdot(\vec E\times\vec B)+{\cal E}^2\,\vec B^2+\vec E^2\,\vec P^2\\
   &-&(\vec E\cdot\vec P)^2-(\vec E\cdot\vec B)^2-(\vec P\cdot\vec B)^2\\
   &=&({\cal E}\,\vec B+\vec E\times\vec P)^2-(\vec E\cdot\vec B)^2-(\vec P\cdot\vec B)^2\\
\omega_1 &=&\sqrt{K_1+2\,\sqrt{K_2}}\\
\omega_2 &=&\sqrt{K_1-2\,\sqrt{K_2}}\\
\mathrm{Det}({\bf F})&=&K_1^2-4\,K_2\\
\label{eq_eigenfreq}
\endary
Force matrices of stable systems have purely imaginary eigenvalues~\cite{Arnold}, so
that for stable systems one has $K_2>0$ and $K_1>2\,\sqrt{K_2}$.

Using the notation of the EMEQ a general symplex ${\bf F}$ is given explicitely by
{\small\begary{rcl}
{\bf F}&=&\bmtx{cccc}
-E_x&E_z+B_y&E_y-B_z&B_x\\
E_z-B_y&E_x&-B_x&-E_y-B_z\\
E_y+B_z&B_x&E_x&E_z-B_y\\
-B_x&-E_y+B_z&E_z+B_y&-E_x\\
\emtx\\
&+&\bmtx{cccc}
-P_z&{\cal E}-P_x&0&P_y\\
-{\cal E}-P_x&P_z&P_y&0\\
0&P_y&-P_z&{\cal E}+P_x\\
P_y&0&-{\cal E}+P_x&P_z\\
\emtx\,,
\label{eq_edeq1}
\endary}
Note that ${\bf F}$ is block-diagonal, if $B_x=B_z=E_y=P_y=0$.

%%%%%%%%%%%%%%%%%%%%%%%%%%%%%%%%%%%%%%%%%%%%%%%%%%%%%%%%%%%%%%%%%%%%%%%%%%%%%%%%
\section{Decoupling of 2-dimensional systems}
\label{sec_decoupling_irregular}
%%%%%%%%%%%%%%%%%%%%%%%%%%%%%%%%%%%%%%%%%%%%%%%%%%%%%%%%%%%%%%%%%%%%%%%%%%%%%%%%

\subsection{The geometrical approach}

In the following we describe a geometrical approach of decoupling that is inspired
by the observation, that in the decoupled force matrix, the scalar products
$\vec E\cdot\vec B$ and $\vec P\cdot\vec B$ vanish~\cite{rdm_paper}. In Hamiltonian form
(see Eqn.~\ref{eq_Hform} below), also the product $\vec P\cdot\vec E$ is zero 
and only the components ${\cal E}$, $P_x$, $E_z$ and $B_y$ remain.
It is therefore instructive to analyze the symplectic transformation properties
of these scalar products. The product $\vec E\cdot\vec B$ is known to be invariant under 
rotations and Lorentz boosts. Formally it is a pseudo-scalar in contrast to the scalar
component representing the mass. Hence one might loosely speak of ``mass components''
and use the abbreviations:
\begary{rcl}
M_r&=&\vec E\cdot\vec B\\
M_g&=&\vec B\cdot\vec P\\
M_b&=&\vec E\cdot\vec P\\
\label{eq_aux_masses}
\endary
The ``mass components'' are invariant under spatial rotations. We may therefore 
proceed with phase rotations and boosts. We introduce the following auxiliary vectors: 
\begary{rcl}
\vec r&\equiv&{\cal E}\,\vec P+\vec B\times \vec E \\
\vec g&\equiv&{\cal E}\,\vec E+\vec P\times \vec B \\
\vec b&\equiv&{\cal E}\,\vec B+\vec E\times \vec P \,,
\label{eq_aux_vecs}
\endary
so that $K_2$ from Eqn.~\ref{eq_eigenfreq} can be written as
\begeq
K_2=\vec b^2-M_r^2-M_g^2\,.
\endeq
It is easy to see that ${\vec g}$, ${\vec r}$ and ${\vec b}$ transform under spatial 
rotations just like usual vectors. It is also quite obvious that the vector $\vec g$
equals the usual Lorentz force and the vector $\vec b$ equals the ``Lorentz force'' 
of a particle with magnetic charge, as the role of $\vec E$ and $\vec B$ is exchanged 
compared to $\vec g$ in the algebraic way that corresponds to a duality rotation through
an angle of ${\pi\over 2}$~\cite{rdm_paper}. 

One finds the following products:
\begary{rcl}
\vec g^2&=&-2\,{\cal E}\,\vec P\cdot(\vec E\times\vec B)+{\cal E}^2\,\vec E^2+\vec B^2\,\vec P^2-M_g^2\\
\vec r^2&=&-2\,{\cal E}\,\vec P\cdot(\vec E\times\vec B)+{\cal E}^2\,\vec P^2+\vec B^2\,\vec E^2-M_r^2\\
\vec b^2&=&-2\,{\cal E}\,\vec P\cdot(\vec E\times\vec B)+{\cal E}^2\,\vec B^2+\vec E^2\,\vec P^2-M_b^2\\
\vec g\cdot\vec r&=&({\cal E}^2-\vec B^2)\,M_b+M_r\,M_g\\
\vec g\cdot\vec b&=&({\cal E}^2-\vec P^2)\,M_r+M_g\,M_b\\
\vec r\cdot\vec b&=&({\cal E}^2-\vec B^2)\,M_g+M_r\,M_b\\
\endary
We introduce the following abbreviations for a better readability
\begary{rclp{5mm}rcl}
c&=&\cos{(\eps)} && s&=&\sin{(\eps)}\\
c_2&=&\cos{(2\,\eps)} && s_2&=&\sin{(2\,\eps)}\\
C&=&\cosh{(\eps)}&&S&=&\sinh{(\eps)}\\
C_2&=&\cosh{(2\,\eps)}&&S_2&=&\sinh{(2\,\eps)}\\
\label{eq_aux_angles}
\endary
The phase rotation generated by $\y_0$ yields:
\begary{rcl}
{\vec g}'&=&\vec g\,c+\vec r\,s\\
{\vec r}'&=&\vec r\,c-\vec g\,s\\
{\vec b}'&=&\vec b\\
\endary
The transformation of the mass components is listed in Tab.~\ref{tab_mass_trans}.
\begin{table}
\begin{tabular}{|c|c|c|c|}\hline
             & $M_r'$                     & $M_g'$                 & $M_b'$           \\\hline\hline
  $\y_0$     & $M_r\,c+M_g\,s$            & $M_g\,c-M_r\,s$        & $M_b\,c_2+{\vec P^2-\vec E^2\over 2}\,s_2$\\\hline
  $\y_1$     & $M_r\,C-(\vec b)_x\,S$     & $M_g$                  & $M_b\,C-(\vec r)_x\,S$\\\hline
  $\y_2$     & $M_r\,C-(\vec b)_y\,S$     & $M_g$                  & $M_b\,C-(\vec r)_y\,S$\\\hline
  $\y_3$     & $M_r\,C-(\vec b)_z\,S$     & $M_g$                  & $M_b\,C-(\vec r)_z\,S$\\\hline
  $\y_4$     & $M_r$                      & $M_g\,C+(\vec b)_x\,S$ & $M_b\,C+(\vec g)_x\,S$\\\hline
  $\y_5$     & $M_r$                      & $M_g\,C+(\vec b)_y\,S$ & $M_b\,C+(\vec g)_y\,S$\\\hline
  $\y_6$     & $M_r$                      & $M_g\,C+(\vec b)_z\,S$ & $M_b\,C+(\vec g)_z\,S$\\\hline
\end{tabular}
\caption{Table of transformed ``mass components'' for symplectic transformations in 2 dimensions.
Compare Eqns.~\ref{eq_aux_masses},~\ref{eq_aux_vecs} and~\ref{eq_aux_angles}. 
\label{tab_mass_trans}}
\end{table}
From the discussion of the normal form of the force matrix in Ref.~\cite{rdm_paper} it
follows, that decoupling to block-diagonal form is done by a transformation that makes
$P_y=E_y=B_x=B_z=0$. {\it Geometrically} this means, that $\vec B$ has to be aligned along 
the y-axis and the vectors $\vec P$ and $\vec E$ should be in the plane perpendicular to $\vec B$.
In a first step, the decoupling of a two-dimensional harmonic oscillator
requires the (partial) orthogonalization of the (3-dimensional) ``vectors'' $\vec E$, $\vec B$ and $\vec P$:
\begary{rcl}
M_r&=&\vec E\cdot\vec B\to 0\\
M_g&=&\vec P\cdot\vec B\to 0\,,
\endary
which can be interpreted as a geometrical ``decoupling''. The alignment of $\vec B$ along the y-axis
in a second step is simple. A transformation to what we call ``Hamiltonian'' form
\begeq
{\bf F}_d=\bmtx{cccc}
0&\alpha&0&0\\
-\beta&0&0&0\\
0&0&0&\y\\
0&0&-\delta&0\emtx\,,
\label{eq_Hform}
\endeq
requires additionally to make $E_x=P_z=0$, which can again by done in two steps, orthogonalization
\begeq
M_b=\vec E\cdot\vec P\to 0\,,
\endeq
and subsequent alignment of $\vec E$ and $\vec P$.
The general form of symplectic transformations has been described in some detail in Ref.~\cite{rdm_paper}, 
here we give only a brief summary. A symplectic transformation matrix ${\bf R}_b$ is generated 
by a basic symplex $\y_b$ with $b\in [0\dots 9]$ and controlled by a parameter $\eps$:
\begary{rcl}
{\bf R}_b&=&\exp{(\y_b\,{\eps\over 2})}\\
{\bf R}_b^{-1}&=&\exp{(-\y_b\,{\eps\over 2})}\\
{\bf F}&\to&{\bf R}_b\,{\bf F}\,{\bf R}_b^{-1}\\
\endary
The effect of a basic symplex $\y_b$ depends on its ``signature'', which
is positive for symmetric and negative for skew-symmetric $\y_b$:
\begary{rcl}
{\bf R}_b&=&\left\{
\begin{array}{lcl}
{\bf 1}\,\cos{(\eps/2)}+\y_b\,\sin{(\eps/2)}&\mathrm{for}&\y_b^2=-{\bf 1}\\
{\bf 1}\,\cosh{(\eps/2)}+\y_b\,\sinh{(\eps/2)}&\mathrm{for}&\y_b^2={\bf 1}\\
\end{array}\right.\,,
\label{eq_Rb}
\endary
where the bold printed ${\bf 1}$ is the unity matrix. Note that transformations
with $\y_b^2=-{\bf 1}$ ($+{\bf 1}$) are called rotations (boosts), respectively.
Explicitely, $\y_0$ is the generator of a ``phase rotation'', $\y_b\,\,\,b\in[7,8,9]$
are ``spatial rotations`` with respect to the $x$, $y$ and $z$-axis and
$\y_b\,\,\,b\in[4,5,6]$ generate ``Lorentz boosts'' with respect to the $x$, $y$ and $z$-axis.
The ``phase boosts'' generated by $\y_b\,\,\,b\in[1,2,3]$ are combinations of
phase rotations and Lorentz boosts.
The parameter $\eps$ is called ``angle'' in case of rotations and ``rapidity''
in case of boosts.
As the decoupling requires a sequence of transformations, we emphasize that
the RDM-coefficients have to be updated according to Eq.~\ref{eq_rdmcoeffs} after
each transformation.

Inspection of Tab.~\ref{tab_mass_trans} shows that a straightforward strategy is the following:
\begin{itemize}
\item $M_g\to 0$: Make a phase rotation generated by $\y_0$ with angle $\eps=\arctan{({M_g\over M_r})}$.
This will always work independent on the size of $M_i$.
\item $\vec b\to \vert\vec b\vert\,\vec e_y$: Align the vector $\vec b$ along the $y$-axis by the 
spatial rotations with ${\bf R}_7$ and an angle of $\eps=\arctan{({b_z\over b_y})}$ and with ${\bf R}_9$ 
through an angle of $\eps=-\arctan{({b_x\over b_y})}$. Such rotations can always be done.
\item $M_r\to 0$: Boost using $\y_2$ and angle $\eps=\mathrm{arctanh}{({M_r\over b_y})}$.
\end{itemize}
The last transformation is only possible, if $\vert M_r\vert<\vert b_y\vert=\vert\vec b\vert$:
\begary{rcl}
(\vec E\cdot\vec B)^2&\le&-2\,{\cal E}\,\vec P\cdot(\vec E\times\vec B)+{\cal E}^2\,\vec B^2+\vec E^2\,\vec P^2-(\vec E\cdot\vec P)^2\\
\label{eq_transcond}
\endary
The first transformations result in $\vec P\cdot\vec B=0$, so that Eq.~\ref{eq_transcond}
is identical to the requirement that $K_2\ge 0$ (see Eq.~\ref{eq_eigenfreq}).
This means that the eigenvalues are either located on the real or imaginary axis, but not 
off-axis in the complex plane. If this condition is fulfilled, then the vector-components 
$(\vec g)_y$ and $(\vec r)_y$, $(\vec b)_x$ and $(\vec b)_z$ are zero after the decoupling 
transformations have been applied. It follows from $M_r=\vec E\cdot\vec B=0$ and
$M_g=\vec P\cdot\vec B=0$ and Eq.~\ref{eq_aux_vecs} that $\vec E\cdot\vec b=0$ and
$\vec P\cdot\vec b=0$, and since we aligned $\vec b$ along the $y$-axis, we have $E_y=0$ and
$P_y=0$, so that with $\vec b$ also $\vec B$ is aligned along the $y$-axis and $B_x=B_z=0$.
If we compare this with Eq.~\ref{eq_edeq1}, then we note that the matrix ${\bf F}$ is now
block-diagonal.

That is: we found a symplectic decoupling algorithm for both - systems with purely imaginary 
eigenvalues, which are called ``strongly'' stable~\cite{Arnold}, 
and unfocused systems with purely real eigenvalues. That the algorithm works in both cases
equally well, is important for instance in the case of transverse-longitudinal coupling 
with space charge in cyclotrons~\cite{cyc_paper}.

We continue the discussion of force matrices with eigenvalues off axis in the complex plane
in Sec.~\ref{sec_complex} and assume for now, that $K_2>0$.
Using the abbreviations
\begary{rcl}
M_x&=&\sqrt{M_r^2+M_g^2}\\
b_{yz}&=&\sqrt{b_y^2+b_z^2}\,,
\endary
the RDM-coefficients of the block-diagonal (decoupled) force matrix are given by:
\begary{rcl}
{\cal E}'&=&{\cal E}\,\sqrt{1-{M_x^2\over\vec b^2}}\\
P_x'&=&{P_x\,M_r-E_x\,M_g\over M_x}\,{\sqrt{\vec b^2-M_x^2}\over b_{yz}}\\
P_z'&=&{\sqrt{\vec b^2-M_x^2}\over\vec b^2\,M_x\,b_{yz}}\,\left[M_g\,(b_z\,E_y-b_y\,E_z)+M_r\,(b_y\,P_z-b_z\,P_y)\right]\\
E_x'&=&{\vec b^2\,(M_r\,E_x+M_g\,P_x)-{\cal E}\,b_x\,M_x^2\over M_x\,b_{yz}\,\vert b\vert}\\
E_z'&=&{M_r\,(b_y\,E_z-b_z\,E_y)+M_g\,(b_y\,P_z-b_z\,P_y)\over M_x\,b_{yz}}\\
B_y'&=&{{\cal E}\,\vec B^2-\vec P\cdot(\vec E\times\vec B)\over\vert\vec b\vert}\\
B_x'&=&B_z'=E_y'=P_y'=0\\
\endary

In order to bring the block-diagonal force matrix to Hamiltonian form, one may apply the
following transformations:
\begin{itemize}
\item $M_b\to 0$: Use another phase rotation with $\y_0$ with $\eps={1\over 2}\,\arctan{({2\,M_b\over \vec E^2-\vec P^2})}$
\item $P_z\to 0$: Use rotation about $y$-axis with $\y_8$ with $\eps=-\arctan{({P_z\over P_x})}$.
\end{itemize}
After these two rotations, the matrix has Hamiltonian form, if $K_2>0$ holds. In charged 
particle optics this is usually the case and therefore we consider this method as a generally applicable
decoupling algorithm. 

%%%%%%%%%%%%%%%%%%%%%%%%%%%%%%%%%%%%%%%%%%%%%%%%%%%%%%%%%%%%%%%%%%%%%%%%%%%%%%%%

\subsection{Complex Eigenvalues}
\label{sec_complex}

Even though the problem of complex eigenvalues has not yet been solved for the general
$2\,n\times 2\,n$ case, it is possible to give a solution for the $4\times 4$ case as
we are going to describe here. The more general case of arbitrary $2\,n\times 2\,n$-symplices
with arbitrary (complex) eigenvalues can presumably be solved by a block-diagonalization
with $4\times 4$-blocks for each set of complex conjugate eigenvalues and $2\times 2$-blocks
for each pair of real or imaginary eigenvalues.

If $K_2<0$ the eigenvalues are complex and a block-diagonalization with $2\times 2$-blocks 
is not possible (within the reals). However a simplification of the matrix is possible with 
the aim, that the RDM-coefficients of the transformed matrix have the following structure:
\begary{rclp{5mm}rclp{5mm}rcl}
P_x&=&0&&P_y&=&0&&P_z&=&0\\
E_x&=&0&&B_x&=&0&&B_z&=&0\\
E_z&\ne&0&&E_y&\ne&0&&B_y&\ne&0\\
{\cal E}&=&0&&M_g&=&0&&M_b&=&0\,,
\label{eq_cplx}
\endary
so that one finds $\vec g=0$ and $\vec b=0$ and the auxiliary vector $\vec r$ has
only a single non-vanishing component $r_x$.
We distinguish two cases, the first with ${\cal E}^2<\mathrm{Max}(\vec P^2,\vec E^2)$ and
the second with ${\cal E}^2>\mathrm{Min}(\vec P^2,\vec E^2)$.
In both cases the goal is to let ``energy'' and ``momentum'' vanish by appropriate
Lorentz or phase boosts. Then one may align $\vec B$ along the $y$-axis and rotate
about the $y$-axis to make $E_x=0$. Then the conditions of Eqn.~\ref{eq_cplx} are fulfilled.

\subsubsection{The Low Energy Case}

The decoupling strategy for the first case, i.e. for ${\cal E}^2<\mathrm{Max}(\vec P^2,\vec E^2)$:
\begin{itemize}
\item $M_g\to 0$: Apply a phase rotation ${\bf R}_0$ with angle $\eps_1=\arctan{({M_g\over M_r})}$.
Note that this maximizes $M_r=\vec E\cdot\vec B$.
\item $\vec E\to \vert\vec E\vert\,\vec e_y$: Align the vector $\vec E$ along the $y$-axis by the 
spatial rotations with ${\bf R}_7$ and an angle of $\eps_2=\arctan{({E_z\over E_y})}$ and (after
an update of the RDM-coefficients and a recomputation of the auxiliary vector and mass components) with ${\bf R}_9$ 
about an angle of $\eps_3=-\arctan{({E_x\over E_y})}$. 
\item ${\cal E}\to 0$: Boost using ${\bf R}_2$ and rapidity $\eps_4=\mathrm{arctanh}{({{\cal E}\over E_y})}$.
  According to the assumptions, this is possible and does not change $E_x=0$ or $E_z=0$.
\item $P_x\to 0$: Boost using ${\bf R}_3$ and rapidity $\eps_5=-\mathrm{arctanh}{({P_x\over B_y})}$.
\item $P_z\to 0$: Boost using ${\bf R}_1$ and rapidity $\eps_6=\mathrm{arctanh}{({P_z\over B_y})}$.
  Since ${\cal E}=E_z=E_x=0$, the energy ${\cal E}$ as well as $\vec E$ are unchanged by the boost.
\item $\vec B\to \vert\vec B\vert\,\vec e_y$: Align the vector $\vec B$ along the $y$-axis by the 
spatial rotations with ${\bf R}_7$ and an angle of $\eps_7=\arctan{({B_z\over B_y})}$ and (after
an update of the RDM-coefficients and a recomputation of the auxiliary vector and mass components) with ${\bf R}_9$ 
about an angle of $\eps_8=-\arctan{({B_x\over B_y})}$. 
\item $E_x\to 0$: Rotate about the $y$-axis with ${\bf R}_8$ with an angle of $\eps_9=\arctan{({E_x\over E_z})}$.
\end{itemize}

\subsubsection{The Intermediate Energy Case}

The case where ${\cal E}^2>\mathrm{Min}(\vec P^2,\vec E^2)$ but $K_2<0$ might be called ``intermediate'', since
the energy is large compared to the ``low energy'' case, but not large enough to make $K_2>0$. The following
procedure leads to the state described by Eqn.~\ref{eq_cplx}:
\begin{itemize}
\item $M_b\to 0$: Apply a phase rotation ${\bf R}_0$ with angle $\eps_1={1\over 2}\,\arctan{({2\,M_b\over\vec E^2-\vec P^2})}$.
Note that this transformation minimizes $\vec P^2$.
\item $\vec P\to \vert\vec P\vert\,\vec e_y$: Align the vector $\vec P$ along the $y$-axis by the 
spatial rotations with ${\bf R}_7$ and an angle of $\eps_2=\arctan{({P_z\over P_y})}$ and (after
an update of the RDM-coefficients and a recomputation of the auxiliary vector and mass components) with ${\bf R}_9$ 
about an angle of $\eps_3=-\arctan{({P_x\over P_y})}$. Since $M_b=\vec E\cdot\vec P=0$ one also has now $E_y=0$.  
\item $P_y\to 0$: Lorentz boost using ${\bf R}_5$ and rapidity $\eps_4=\mathrm{arctanh}{({P_y\over {\cal E}})}$.
\item $\vec B\to \vert\vec B\vert\,\vec e_y$: Align the vector $\vec B$ along the $y$-axis by the 
spatial rotations with ${\bf R}_7$ and an angle of $\eps_6=\arctan{({B_z\over B_y})}$ and (after
an update of the RDM-coefficients and a recomputation of the auxiliary vector and mass components) with ${\bf R}_9$ 
about an angle of $\eps_7=-\arctan{({B_x\over B_y})}$. 
\item ${\cal E}\to 0$: Boost using ${\bf R}_2$ and rapidity $\eps_8=\mathrm{arctanh}{({{\cal E}\over E_y})}$.
\item $E_x\to 0$: Rotate about the $y$-axis with ${\bf R}_8$ with an angle of $\eps_9=\arctan{({E_x\over E_z})}$.
\end{itemize}

In both cases the transformed matrix ${\bf F}$ then has the form
\begeq
{\bf F}=E_y\,\y_5+E_z\,\y_6+B_y\,\y_8\,.
\endeq
In order to bring it to Hamiltonian form, one applies the transformation ${\bf R}_2$ with
an ``angle'' of $i\,\pi/2$:
\begeq
{\bf R}=\exp{(i\,{\pi/4}\,\y_2)}={1\over\sqrt{2}}\,({\bf 1}+i\,\y_2)\,.
\endeq
so that ${\bf F}\to {\bf R}\,{\bf F}\,{\bf R}^{-1}$ is:
\begeq
{\bf F}=-i\,E_y\,\y_0+E_z\,\y_6+B_y\,\y_8\,.
\endeq
Note that the complex eigenvalues of a force matrix with $K_2<0$ all lie on a circle of radius 
$\rho=(K_1^2+4\,\vert K_2\vert)^{1/4}$ in the complex plane.

%%%%%%%%%%%%%%%%%%%%%%%%%%%%%%%%%%%%%%%%%%%%%%%%%%%%%%%%%%%%%%%%%%%%%%%%%%%%%%%%

\subsection{Decoupling n-dimensional Symplices}
\label{sec_2nx2n}

The general eigenvalue problem of symplices (Hamiltonian matrices) is an area of intense 
research. The algorithm presented above is based on a physical and geometrical analysis of 2-dimensional
linear symplectic systems. As described before, the algorithm is limited to symplices that
have real or imaginary eigenvalues, but a generalization to include complex eigenvalues 
might be possible - even though not urgently required in charged particle optics~\footnote{The ${\bf S}$-matrix
for instance will never have complex eigenvalues as it is derived from the matrix of second moments.
Complex eigenvalues would only be possible for correlations with a modulus greater than one.}.

In order to decouple symplectic systems with more than two degrees of freedom, the described algorithm 
can be used in an iterative scheme analogous to the Jacobi method for symmetric matrices\footnote{
Jacobi introduced a method to iteratively diagonalize real symmetric matrices by a sequence of 
orthogonal transformations each of which diagonalizes a $2\times 2$ submatrix~\cite{Jacobi}.}.
If all eigenvalues are real or imaginary, it is possible to avoid computations using complex numbers. 
The $2\,n\,\times 2\,n$-symplex is then regarded as a $n\times n$ matrix of $2\times 2$-blocks.
We tested a pivot search that picks the maximum average square amplitude of all non-diagonal blocks 
${\bf B}_{ij}$. The blocks ${\bf B}_{ii}$, ${\bf B}_{ij}$, ${\bf B}_{ji}$ and ${\bf B}_{jj}$ are 
then analyzed as $4\times 4$-symplices and the symplectic similarity transformation that 
block-diagonalizes this submatrix is applied, so that ${\bf\tilde B}_{ij}={\bf\tilde B}_{ji}=0$ holds. 
This iterative scheme allows to compute simultaneously the symplectic transformation matrix and the 
resulting block-diagonal or Hamiltonian form symplex with high precision. 

Given $\{x\}$ is a sequence of random numbers between zero and one, then one may construct random 
symmetric $2\,n\times 2\,n$-matrices ${\bf A}$ according to the rule:
\begeq
{\bf A}_{ij}={\bf A}_{ji}=\left\{\begin{array}{rcl}
x-{1\over 2}&\mathrm{for}&i\neq j\\
n+x&\mathrm{for}&i=j\\
\end{array}\right.
\endeq
The increase of the diagonal terms helps to avoid complex eigenvalues. The symplex to decouple
is then given by ${\bf F}=\y_0\,{\bf A}$. We tested the algorithm with these random matrices 
up to $n=12$ and logged the number of $4\times 4$-diagonalization steps.
\begin{figure}
\includegraphics[width=7.5cm]{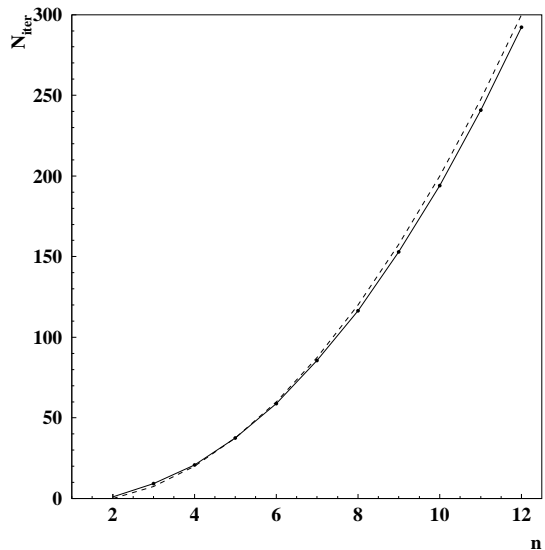}
\caption{
Solid line: Number of iterations required to bring a $2\,n\times 2\,n$ symplex (Hamiltonian matrix) 
to Hamiltonian form. Dashed line: Approximation by $5\,{n\,(n-2)\over 2}$. The number $n_b$ of non-diagonal 
$2\times 2$-blocks is $n_b={n\,(n-1)\over 2}$.
\label{fig_iter}
}
\end{figure}
Fig.~\ref{fig_iter} shows the average number of iterations that is required to compute the 
transformation that brings a $2\,n\times 2\,n$ symplex to Hamiltonian form, i.e. into the form:
\begeq
\bmtx{ccccc}
0&\beta_1&\dots&0&0\\
-\gamma_1&0&\dots&0&0\\
\vdots  &\vdots&&\vdots&\vdots\\
0&0&\dots&0&\beta_n\\
0&0&\dots&-\gamma_n&0\\
\emtx
\endeq

\subsection{Diagonalization}

In order to proceed from Eqn.~\ref{eq_Hform} towards diagonalization, the 
matrix is {\it scaled} using the generators $\y_3$ and $\y_4$:
\begary{rcl}
{\bf R} &=&\exp{[(\y_3+\y_4) {s\over 2}+(\y_3-\y_4) {t\over 2}]}\\
        &=&\mathrm{Diag}(\exp{(-s)},\exp{(s)},\exp{(-t)},\exp{(t)})\\
       s&=&\exp{(\log{(\vert{\alpha\over\beta}\vert)}/4)}\\
       t&=&\exp{(\log{(\vert{\gamma\over\delta}\vert)}/4)}\\
\endary
so that one obtains (for stable systems), what we call ``normal'' form:
\begary{rcl}
{\bf F}&\to&{\bf R}\,{\bf F}\,{\bf R}^{-1}\\
        &=&\bmtx{cccc}
             0&\w_1&0&0\\
             -\w_1&0&0&0\\
             0&0&0&\w_2\\
             0&0&-\w_2&0\\
           \emtx\,,
\label{eq_normalform}
\endary
where the signs of the frequencies $\w_1$ and $\w_2$ can both be positive and negative, 
depending on the signs of $\alpha$, $\beta$, $\gamma$ and $\delta$.
At this stage, all components of $\vec g$ and $\vec r$ as well as $(\vec b)_x$ and 
$(\vec b)_z$ are zero. Only $(\vec b)_y$ is non-zero.
\begary{rcl}
{\bf F}&=&{\bf E}_0\,\mathrm{Diag}(i\,\w_1,-i\,\w_1,i\,\w_2,-i\,\w_2)\,{\bf E}_0^{-1}\\
{\bf E}_0 &=&{1\over 2}\,\bmtx{cccc}
             1-i& -1+i & 0 & 0\\
             1+i & 1+i & 0 & 0\\
             0 & 0 &1-i & -1+i\\
             0 & 0 &1+i &  1+i\\         
           \emtx\\
        &=&{1\over 2}\,\left({\bf 1}-\y_0+i\,\y_3+i\,\y_6\right)\\
{\bf E}_0\,\y_0\,{\bf E}_0^T&=&\y_0\\
{\bf E}_0^{-1}&=&{\bf E}_0^\dag\\
\endary
That is - the last transformation matrix that is required for diagonalization is not
only symplectic - it is also unitary.

\subsection{Example}

A simplified and idealized cyclotron model with space charge was described, 
which served as an example for an irregular system~\cite{cyc_paper,rdm_paper}. 
Without repeating all details, the constant force matrix has the following form: 
\begeq
{\bf F}=\bmtx{cccc}
0&1&0&0\\
-k_x+K_x&0&0&h\\
-h&0&0&{1\over\y^2}\\
0&0&K_z\,\y^2&0\\
\emtx\,,
\label{eq_eqom_sc}
\endeq
The RDM-coefficients are then given by:
\begary{rcl}
{\cal E}&=&{1\over 4}\,\left(1+k_x-K_x+{1\over\y^2}-\y^2\,K_z\right)\\
P_x&=&{1\over 4}\,\left(-1+k_x-K_x+{1\over\y^2}+\y^2\,K_z\right)\\
P_y&=&P_z=0\\
E_x&=&B_x=0\\
E_y&=&B_z=-{h\over 2}\\
E_z&=&{1\over 4}\,\left(1-k_x+K_x+{1\over\y^2}+\y^2\,K_z\right)\\
B_y&=&{1\over 4}\left(1+k_x-K_x-{1\over\y^2}+\y^2\,K_z\right)\\
\endary
From this one finds for the ``mass'' terms and the vectors $\vec g$, $\vec r$ and $\vec b$:
\begary{rcl}
M_r&=&\vec E\cdot\vec B=-{h\over 4}\,(1+K_z\,\y^2)\\
M_g&=&\vec P\cdot\vec B=0\\
M_b&=&\vec E\cdot\vec P=0\\
\vec g&=&(0,{h\over 4}\,(K_z\,\y^2-1),{1+\y^2\,(k_x-K_x)\,K_z\over 4\,\y^2})^T\\
\vec r&=&({k_x-K_x+\y^2\,(K_z-h^2)\over 4\,\y^2},0,0)^T\\
\vec b&=&(0,{K_z+k_x-K_x\over 4},{h\,(\y^2\,K_z-1)\over 4})^T\\
\endary
According to the geometrical approach, the first transformation can be omitted, since the
``mass'' $M_g$ is zero. The second transformation using $\y_7$ aligns $\vec b$ along
the $y$-axis. The second rotation may again be omitted, since the vector $\vec r$ is already
aligned along the $x$-axis.
The last transformation is a phase boost using $\y_2$ and is sufficient to bring ${\bf F}$ 
into block-diagonal form. This transformation would usually change the value of $M_b$, but
here it does not, since $M_b=(\vec r)_y=0$ as can be seen from Tab.~\ref{tab_mass_trans}.
Hence $M_b$ remains zero - $M_g$ is invariant under both transformations. Hence, 
all ``mass terms'' are then zero after the described two transformations so that the 
system is decoupled.

%%%%%%%%%%%%%%%%%%%%%%%%%%%%%%%%%%%%%%%%%%%%%%%%%%%%%%%%%%%%%%%%%%%%%%%%%%%%%%%%
\subsection{Operators, Expectation Values and Lax Pairs}
%%%%%%%%%%%%%%%%%%%%%%%%%%%%%%%%%%%%%%%%%%%%%%%%%%%%%%%%%%%%%%%%%%%%%%%%%%%%%%%%

Coupled linear optics is in its essence (as quantum mechanics) a statistical
theory. Since the reference trajectory is fixed, the coordinates are always 
taken relative to the local reference frame and the geometry is (only) locally 
euclidean.
Even though the starting point is the description of single particle motion,
the orbits of single particles are usually both, hard to access experimentally
and of low practical value. The description of the beam by average values in 
contrast is both - measureable and of high value. The use of symplectic 
transformations leaves the expectation values unchanged. We can therefore
evaluate the expectation values of any operator ${\bf O}$ in an arbitrary reference frame:
\begary{rcl}
\langle{\bf O}\rangle&\equiv&\langle\bar\psi\,{\bf O}\,\psi\,\rangle\\
                     &=&\langle\psi^T\,\y_0\,{\bf R}^{-1}\,{\bf R}\,{\bf O}\,{\bf R}^{-1}\,{\bf R}\,\psi\,\rangle\\
                     &=&\langle\psi^T\,\y_0\,{\bf R}^{-1}\,{\bf \tilde O}\,\tilde\psi\,\rangle\\
                     &=&\langle\psi^T\,{\bf R}^T\,\y_0\,{\bf \tilde O}\,\tilde\psi\,\rangle\\
                     &=&\langle\tilde{\bar\psi}\,{\bf \tilde O}\,\tilde\psi\,\rangle\,,
\endary
since for symplectic ${\bf R}$ we have
\begary{rcl}
{\bf R}^T\,\y_0&=&\y_0\,{\bf R}^{-1}\\
{\bf R}^T\,\y_0\,{\bf R}&=&\y_0\\
\endary
The time derivative of the expectation value of an arbitrary operator ${\bf O}$, that does not 
explicitely depend on time, is:
\begary{rcl}
{d\over d\tau}\left(\bar\psi\,{\bf O}\,\psi\right)&=&\dot{\bar\psi}\,{\bf O}\,\psi+\bar\psi\,{\bf O}\,\dot\psi\\
&=&\psi^T\,{\bf F}^T\,\y_0\,{\bf O}\,\psi+\bar\psi\,{\bf O}\,{\bf F}\,\psi\\
&=&\bar\psi\,({\bf O}\,{\bf F}-{\bf F}\,{\bf O})\,\psi\\
\label{eq_OpEx}
\endary
Equations of the form (here ${\bf S}=\sigma\,\y_0$)
\begeq
{\bf\dot S}={\bf F}\,{\bf S}-{\bf S}\,{\bf F}\,,
\label{eq_eqomS}
\endeq
appear frequently in the theory of coupled linear optics and it is worth
mentioning that Eq.~\ref{eq_eqomS} is a so-called {\it Lax representation} and
the operators ${\bf S}$ and ${\bf F}$ are a so-called {\it Lax pair}~\cite{Lax,Lax2}.
As a consequence, the expressions
\begeq
I_k=Tr({\bf S}^k)
\label{eq_lax}
\endeq
are first integrals of motion, where $Tr()$ is the trace.
Using again the EMEQ to express the elements of ${\bf S}$, one finds:
\begary{rcl}
I_1&=&Tr({\bf S})=0\\
I_2&=&Tr({\bf S}^2)=-4\,\left({\cal E}^2-\vec P^2+\vec B^2-\vec E^2\right)=-4\,K_1\\
I_3&=&Tr({\bf S}^3)=0\\
I_4&=&Tr({\bf S}^4)=4\,(K_1^2+4\,K_2)\\
\endary
The values of $K_1$ and $K_2$ are (as expected) first integrals and constants of motion. 
The complete expression for ${\bf S}^4$ is
\begeq
{\bf S}^4=(K_1^2+4\,K_2)\,{\bf 1}-4\,K_1\,\left(M_g\,\y_{10}+M_r\,\y_{14}+\vec b\,\y_{14}\,\vec\y\right)\,.
\endeq
Another derivation of Eqn.~\ref{eq_lax} has been given in~\cite{DNR}.

\section{Summary and Outlook}

A powerful method for symplectic decoupling of the n-dimensional non-dissipative harmonic
oscillator has been developed. The method apparently is stable, of the order ${\cal O}(n^2)$
and works with purely real or purely imaginary eigenvalues, for which a Hamiltonian Schur form 
does not always exists~\cite{CM}.
The resulting block-diagonal symplex can be used to compute the $\sigma$-matrix
of matched beam ellipsoids of linear coupled systems in charged particle 
optics~\cite{Wolski,rdm_paper,cyc_paper}. Another application is the production
of multivariate gaussian distributions for a given covariance matrix~\cite{stat_paper}.

The presented parametrization gives deep insight into the general nature of
coupling and might be instructive also in other areas of physics. The algebraic problem
of finding the eigenvalues and eigenvectors of a two-dimensional symplectic system
was solved using geometrical arguments based on the use of the real Dirac matrices
and the electromechanical equivalence.

\begin{acknowledgments}
We would like to mention the work of D. Hestenes, who emphasised
the geometrical significance of the Dirac algebra that he called 
{\it space-time algebra}~\cite{Hestenes}. The idea to introduce 
the EMEQ is inspired by his work.

Mathematica\textsuperscript{\textregistered} has been used for some of the
symbolic calculations. Additional software has been written in ``C'' and been compiled 
with the GNU\textsuperscript{\copyright}-C++ compiler 3.4.6 on Scientific Linux.
The CERN library (PAW) was used to generate the figure.
\end{acknowledgments}

\begin{appendix}

\section{The $\y$-Matrices}
\label{sec_app1}
%%%%%%%%%%%%%%%%%%%%%%%%%%%%%%%%%%%%%%%%%%%%%%%%%%%%%%%%%%%%%%%%%%%%%%%%%%%%%%%%

To complete the list of the real $\y$-matrices used throughout this paper:
{\small
\begeq
\begin{array}{rclp{4mm}rcl}
\y_0&=&\bmtx{cccc}
   0 &   1  &  0 &   0\\
  -1 &   0  &  0 &   0\\
   0 &   0  &  0 &   1\\
   0 &   0  & -1 &   0\\
\emtx&&
 \y_1&=&\bmtx{cccc}
   0 &  -1  &  0 &   0\\
  -1 &   0  &  0 &   0\\
   0 &   0  &  0 &   1\\
   0 &   0  &  1 &   0\\
\emtx\\
 \y_2&=&\bmtx{cccc}
   0 &   0  &  0 &   1\\
   0 &   0  &  1 &   0\\
   0 &   1  &  0 &   0\\
   1 &   0  &  0 &   0\\
\emtx&&
 \y_3&=&\bmtx{cccc}
  -1 &   0  &  0 &   0\\
   0 &   1  &  0 &   0\\
   0 &   0  & -1 &   0\\
   0 &   0  &  0 &   1\\
\emtx\\
\y_{14}&=&\y_0\,\y_1\,\y_2\,\y_3;&&\y_{15}&=&{\bf 1}\\
\y_4&=&\y_0\,\y_1;&&\y_7&=&\y_{14}\,\y_0\,\y_1=\y_2\,\y_3\\
\y_5&=&\y_0\,\y_2;&&\y_8&=&\y_{14}\,\y_0\,\y_2=\y_3\,\y_1\\
\y_6&=&\y_0\,\y_3;&&\y_9&=&\y_{14}\,\y_0\,\y_3=\y_1\,\y_2\\
\y_{10}&=&\y_{14}\,\y_0&=&\y_1\,\y_2\,\y_3&&\\
\y_{11}&=&\y_{14}\,\y_1&=&\y_0\,\y_2\,\y_3&&\\
\y_{12}&=&\y_{14}\,\y_2&=&\y_0\,\y_3\,\y_1&&\\
\y_{13}&=&\y_{14}\,\y_3&=&\y_0\,\y_1\,\y_2&&\\
\end{array}
\endeq
}

\section{Floquet Theorem}
\label{app_floquet}

If the matrix ${\bf A}$ in Eqn.~(\ref{eq_Hamiltonian}) and hence the forces are not constant, 
but periodic (${\bf F}(t+T)={\bf F}(t)$), then Floquet's theorem can be applied 
and the solution has the general form~\cite{Talman,MHO}:
\begeq
{\bf M}(t)={\bf K}(t)\,\exp{({\bf\bar F}\,t)}\,,
\label{eq_floquet_theorem}
\endeq
where ${\bf K}(t)$ is symplectic and periodic with period $T$.
\begary{rcl}
{\bf M}(0)&=&{\bf 1}\,\,\Rightarrow\,\,{\bf K}(0)={\bf 1}\\
{\bf K}(t+T)&=&{\bf K}(t)\,\,\Rightarrow\,\,{\bf K}(T)={\bf 1}\\
\endary
The transfer matrix of one period of length $T$ (``one-turn-transfer-matrix'')
\begeq
{\bf M}(T)={\bf M}_T=\exp{({\bf\bar F}\,T)}
\endeq
is identical to the transfer matrix for a system with the constant force matrix 
${\bf\bar F}$ and length $T$. {\it In this sense} ${\bf\bar F}$ is the ``average'' 
or ``effective'' force matrix with respect to one turn and can formally be 
written as~\cite{Talman}:
\begeq
{\bf\bar F}={1\over T}\,\ln{({\bf M}_T)}\,.
\label{eq_force_aver}
\endeq
From Eqs.~\ref{eq_floquet_theorem} one derives in a few steps~\cite{Talman,Leach}:
\begeq
{\bf\dot K}={\bf F}\,{\bf K}-{\bf K}\,{\bf\bar F}\,.
\endeq
If the canonical transformation represented by ${\bf K}$ has been applied to the state 
vector, then with ${\bf K}(0)={\bf 1}$ it follows:
\begeq
\tilde\psi(t)={\bf K}^{-1}\,\psi(t)=\exp{({\bf\bar F}\,t)}\,\tilde\psi(0)={\bf\tilde M}\,\tilde\psi(0)\,.
\endeq
Note that the knowledge of ${\bf K}$ is not required to solve the decoupling problem,
as long as the one-turn-transfer matrix ${\bf M}_T$ is known. ${\bf M}_T$ can
either be obtained as a product of the transfer matrices of all beamline elements
or simply by numerical integration. If the matched beam distribution has been found at an
arbitrary (known) position $s=0$ along the closed reference orbit, then the matched distribution  
can be computed for any position $s$ using:
\begeq
\sigma(s)={\bf M}(s)\,\sigma(0)\,{\bf M}^T(s)\,.
\endeq

\section{Quick Guide to Decoupling}

To start with it is required to have either the average or constant force matrix 
${\bf F}$ or the symplectic transfer matrix ${\bf M}$ that represents a complete 
turn or (cyclotron) sector. In the latter case one computes an auxiliary force 
matrix by
\begeq
{\bf M}_s={1\over 2}\,({\bf M}+\y_0\,{\bf M}^T\,\y_0)\,,
\endeq
while the usual (effective) force matrix has the form
\begeq
{\bf F}={\bf E}\,\mathrm{Diag}(i\,\w_1,-i\,\w_1,i\,\w_2,-i\,\w_2)\,{\bf E}^{-1}\,,
\endeq
$\w_i$ being the betatron frequencies. The auxiliary matrix has the same structure
\begeq
{\bf M}_s={\bf E}\,\mathrm{Diag}(i\,s_1,-i\,s_1,i\,s_2,-i\,s_2)\,{\bf E}^{-1}\,,
\endeq
but different eigenvalues $s_i=\sin{(\w_i\,\tau)}$, where $\w_i\,\tau=2\pi\,Q_i$ 
with the betatron tunes $Q_i$. 

Now compute the RDM-coefficients according to:
\begary{rcl}
{\cal E}&=&-Tr({\bf F}\,\y_0+\y_0\,{\bf F})/8\\
P_x&=&Tr({\bf F}\,\y_1+\y_1\,{\bf F})/8\\
P_y&=&Tr({\bf F}\,\y_2+\y_2\,{\bf F})/8\\
P_z&=&Tr({\bf F}\,\y_3+\y_3\,{\bf F})/8\\
E_x&=&Tr({\bf F}\,\y_4+\y_4\,{\bf F})/8\\
E_y&=&Tr({\bf F}\,\y_5+\y_5\,{\bf F})/8\\
E_z&=&Tr({\bf F}\,\y_6+\y_6\,{\bf F})/8\\
B_x&=&-Tr({\bf F}\,\y_7+\y_7\,{\bf F})/8\\
B_y&=&-Tr({\bf F}\,\y_8+\y_8\,{\bf F})/8\\
B_z&=&-Tr({\bf F}\,\y_9+\y_9\,{\bf F})/8\\
\endary
Note that the coefficients for $\y_k$ with $k\in [10,\dots, 15]$ must be zero 
- otherwise the system is not symplectic.
Then compute the eigenvalues and auxiliary vectors $\vec r,\vec g,\vec b$ 
according to Eq.~\ref{eq_eigenfreq}, \ref{eq_aux_masses} and \ref{eq_aux_vecs}.
Construct the transformation matrices ${\bf R}_b$ according to:
\begary{rcl}
{\bf R}_b&=&\left\{\begin{array}{lcr}
{\bf 1}\,\cos{(\eps/2)}+\y_b\,\sin{(\eps/2)}&\mathrm{for}&b\in [0,7,8,9]\\
{\bf 1}\,\cosh{(\eps/2)}+\y_b\,\sinh{(\eps/2)}&\mathrm{for}&b\in [1,\dots,6]\\
\end{array}\right.\\
{\bf R}_b^{-1}&=&\left\{\begin{array}{lcr}
{\bf 1}\,\cos{(\eps/2)}-\y_b\,\sin{(\eps/2)}&\mathrm{for}&b\in [0,7,8,9]\\
{\bf 1}\,\cosh{(\eps/2)}-\y_b\,\sinh{(\eps/2)}&\mathrm{for}&b\in [1,\dots,6]\\
\end{array}\right.
\endary
Transform with $\y_0$ and $\eps=\arctan{\left({M_g\over M_r}\right)}$:
\begary{rcl}
{\bf F}\to {\bf R}_0\,{\bf F}\,{\bf R}_0^{-1}\,.
\endary
Recompute RDM-coefficients, then transform using $\y_7$ with
$\eps=\arctan{\left({b_z\over b_y}\right)}$.
Recompute RDM-coefficients, then transform using $\y_9$ with
$\eps=\arctan{\left({b_x\over b_y}\right)}$.
Recompute RDM-coefficients, then transform using $\y_2$ with
$\eps=\mathrm{arctanh}{\left({M_r\over b_y}\right)}$. The (auxiliary) force
matrix should now be block-diagonal. Recompute RDM-coefficients, then 
transform with $\y_0$ and $\eps=\arctan{\left({2\,M_b\over \vec E^2-\vec P^2}\right)}$.
Recompute RDM-coefficients, then transform with $\y_8$ and 
$\eps=-\arctan{\left({P_z\over P_x}\right)}$. Now the (auxiliary) force
matrix should have normal form, so that the frequencies (or their sines)
are given by:
\begary{rcl}
\w_1&=&\sqrt{-F_{1,2}\,F_{2,1}}\\
\w_2&=&\sqrt{-F_{3,4}\,F_{4,3}}\\
\endary
The complete transformation is given by:
\begary{rcl}
{\bf R}^{-1}&=&{\bf R}_0^{-1}\cdot{\bf R}_1^{-1}\dots{\bf R}_n^{-1}\\
{\bf R}&=&{\bf R}_n\cdot{\bf R}_{n-1}\dots{\bf R}_0\\
{\bf F}_d&=&{\bf R}\,{\bf F}\,{\bf R}^{-1}\\
\endary
If the auxiliary matrix has been used, then compute the matrix ${\bf\tilde M}_c$ according to
\begeq
{\bf\tilde M}_c={1\over 2}\,{\bf R}\,({\bf M}-\y_0\,{\bf M}^T\,\y_0)\,{\bf R}^{-1}\,.
\endeq
The cosines of the tunes are then given by:
\begary{rcl}
\cos{(\w_1\,\tau)}+\cos{(\w_2\,\tau)}&=&Tr({\bf\tilde M})/2\\
\cos{(\w_1\,\tau)}-\cos{(\w_2\,\tau)}&=&Tr({\bf\tilde M}\,\y_{12}+\y_{12}\,{\bf\tilde M})/4\\
\endary

\section{The Teng and Edwards Ansatz}
\label{sec_ET}

Assume that we have an even number of DOF, so that a $4\,n\times 4\,n$ symplectic matrix ${\bf R}$ 
can be written in block-form according to~\cite{Teng,EdwardsTeng}:
\begeq
{\bf R}=\bmtx{cc}
{\bf A}&{\bf a}\\
{\bf b}&{\bf B}\\
\emtx
\endeq
where all quadratic submatrices are of size $2\,n\times 2\,n$, then the matrix  ${\bf R}$ is symplectic, if
\begary{rcl}
\y_0&=&\bmtx{cc}
{\bf A}&{\bf a}\\
{\bf b}&{\bf B}\\
\emtx\,\y_0\,\bmtx{cc}
{\bf A}^T&{\bf b}^T\\
{\bf a}^T&{\bf B}^T\\
\emtx\\
&=&\bmtx{cc}
{\bf A}\,\y_0\,{\bf A}^T+{\bf a}\,\y_0\,{\bf a}^T&{\bf A}\,\y_0\,{\bf b}^T+{\bf a}\,\y_0\,{\bf B}^T\\
{\bf b}\,\y_0\,{\bf A}^T+{\bf B}\,\y_0\,{\bf a}^T&{\bf b}\,\y_0\,{\bf b}^T+{\bf B}\,\y_0\,{\bf B}^T
\emtx\\
\endary
which yields:
\begary{rcl}
\y_0&=&{\bf A}\,\y_0\,{\bf A}^T+{\bf a}\,\y_0\,{\bf a}^T\\
\y_0&=&{\bf b}\,\y_0\,{\bf b}^T+{\bf B}\,\y_0\,{\bf B}^T\\
0&=&{\bf A}\,\y_0\,{\bf b}^T+{\bf a}\,\y_0\,{\bf B}^T\,,
\endary
where $\y_0$ has - in dependence of the context - to be taken as $2\,n\times 2\,n$ or $4\,n\times 4\,n$.

If one now assumes that ${\bf A}={\bf B}=C\,{\bf 1}$, then it follows that
\begary{rcl}
\y_0\,(1-C^2)&=&{\bf a}\,\y_0\,{\bf a}^T\\
\y_0\,(1-C^2)&=&{\bf b}\,\y_0\,{\bf b}^T\\
{\bf b}&=&\y_0\,{\bf a}^T\,\y_0\,.
\endary
If one assumes furthermore with Teng and Edwards, that $C=\cos{(\phi)}$, then may define
${\bf a}=\sin{(\phi)}\,{\bf a}_s$ and ${\bf b}=\sin{(\phi)}\,{\bf b}_s$ with symplectic matrizes ${\bf a}_s$
and ${\bf b}_s$, respectively:
\begary{rcl}
\y_0&=&{\bf a}_s\,\y_0\,{\bf a}_s^T\\
\y_0&=&{\bf b}_s\,\y_0\,{\bf b}_s^T\\
\endary

It has been shown in Ref.~\cite{cyc_paper}, that $C=\cos{(\phi)}$ is not the general case, since one might
also choose $C=\cosh{(\phi)}$, ${\bf a}=\sinh{(\phi)}\,{\bf a}_s$ and ${\bf b}=\sinh{(\phi)}\,{\bf b}_s$.
In this case one finds 
\begary{rcl}
-\y_0&=&{\bf a}_s\,\y_0\,{\bf a}_s^T\\
-\y_0&=&{\bf b}_s\,\y_0\,{\bf b}_s^T\,,
\endary
i.e. the matrizes ${\bf a}_s$ and ${\bf b}_s$ can also be antisymplectic (symplectic with multiplier $-1$).
Still the matrix ${\bf R}$ remains symplectic. Hence Teng and Edwards limited their treatment in two ways: 
First, they assumed that $C=\cos{(\psi)}$ such that ${\bf R}$ must be a rotation matrix and secondly, they
considered only the case that ${\bf a}$ and ${\bf b}$ are symplectic.

\section{Cosymplices}

The geometric approach is based on the second order terms, i.e. products of the 
RDM coefficients. It is therefore instructive to see where else these terms appear.
For instance one quickly finds the ``mass'' terms and vectors $\vec g$, $\vec r$, $\vec b$ 
in the following products:
\begary{rcl}
{\bf F}\,{\bf F}&=&-({\cal E}^2-\vec P^2+\vec B^2-\vec E^2)\,{\bf 1}+2\,M_r\,\y_{14}\\
                      &+&2\,M_g\,\y_{10}+2\,\vec b\,\y_{14}\,\vec\y\\
{\bf F}\,\y_0\,{\bf F}&=&(3\,{\cal E}^2-\vec P^2-\vec E^2-\vec B^2)\,\y_0-4\,{\cal E}\,{\bf F}\\
                         &+&2\,\vec r\,\vec\y+2\,\vec g\,\y_0\,\vec\y+2\,\vec b\,\y_{14}\,\y_0\,\vec\y\\
{\bf F}\,\y_{14}\,{\bf F}&=&2\,M_b\,\y_{10}-2\,M_r\,{\bf 1}\\
                         &+&({\cal E}^2-\vec P^2+\vec E^2-\vec B^2)\,\y_{14}+2\,\vec g\,\y_{14}\vec\y\\
{\bf F}\,\y_{10}\,{\bf F}&=&2\,M_b\,\y_{14}-2\,M_g\,{\bf 1}\\
                         &+&({\cal E}^2+\vec P^2-\vec E^2-\vec B^2)\,\y_{10}+2\,\vec r\,\y_{14}\vec\y\\
\label{eq_squares}
\endary
So that in the decoupled and normalized case (see Eq.~\ref{eq_normalform}), these products are:
\begary{rcl}
{\bf F}\,{\bf F}&=&-({\cal E}^2-\vec P^2+\vec B^2-\vec E^2)\,{\bf 1}+2\,(\vec b)_y\,\y_{12}\\
{\bf F}\,\y_0\,{\bf F}&=&(3\,{\cal E}^2-\vec P^2-\vec E^2-\vec B^2)\,\y_0-4\,{\cal E}\,{\bf F}\\
                      &+&2\,(\vec b)_y\,\y_8\\
{\bf F}\,\y_{14}\,{\bf F}&=&({\cal E}^2-\vec P^2+\vec E^2-\vec B^2)\,\y_{14}\\
{\bf F}\,\y_{10}\,{\bf F}&=&({\cal E}^2+\vec P^2-\vec E^2-\vec B^2)\,\y_{10}\\
\endary

\section{Expectation Values (Complement)}

In Ref.~\cite{rdm_paper} it has been shown that the expectation values of the RDMs, $f_k$, 
defined by
\begeq
f_k={1\over 2}\,\bar\psi\,\y_k\,\psi\,,
\endeq
vanish for all cosymplices, i.e. for $\y_k$ with $k\in [10,\dots,15]$.
It was also shown that for all symplices (i.e. $\y_k$ with $k\in [0,\dots,9]$ or linear combinations thereof) 
the expectation values $g_k\equiv \bar\psi(\y_k\,{\bf F}+{\bf F}\,\y_k)\psi$ vanish.
Nevertheless nothing was mentioned about the $g_k$ for $k\in[10,\dots,15]$. 
The complement is given in the following:
\begary{rcl}
g_{10}&=&2\,\left(P_x\,f_7+P_y\,f_8+P_z\,f_9-B_x\,f_1-B_y\,f_2+B_z\,f_3\right)\\
g_{11}&=&2\,\left(-{\cal E}\,f_7+B_x\,f_0+P_z\,f_5+E_y\,f_3-P_y\,f_6-E_z\,f_2\right)\\
g_{12}&=&2\,\left(-{\cal E}\,f_8+B_y\,f_0+P_x\,f_6+E_z\,f_1-P_z\,f_4-E_x\,f_3\right)\\
g_{13}&=&2\,\left(-{\cal E}\,f_9+B_z\,f_0+P_y\,f_4+E_x\,f_2-P_x\,f_5-E_y\,f_1\right)\\
g_{14}&=&2\,\left(E_x\,f_7+E_y\,f_8+E_z\,f_9-B_x\,f_4-B_y\,f_5+B_z\,f_6\right)\\
g_{15}&=&2\,\langle {\bf F}\,\rangle\\
\endary
According to Eq.~\ref{eq_OpEx} the expectation values of the operators $g_k$ are:
\begeq
\dot g_k=\bar\psi\,(\y_k\,{\bf F}^2-{\bf F}^2\,\y_k)\,\psi\,.
\label{eq_dotg}
\endeq
The square of the force matrix is given in Eq.~\ref{eq_squares}.
Now we insert this into Eq.~\ref{eq_dotg}. The scalar part commutes
with all $\y_k$ and hence contributes nothing. Since all commutators of symplices
with cosymplices result in cosymplices, we obtain $\dot g_k=0$ for all symplices.
This had to be expected as for all symplices we had $g_k=0$. 
Hence the remaining terms are:
\begary{rcl}
\dot g_{10}&=&4\,\bar\psi\,(M_r\,\y_0+\vec b\,\y_0\,\vec\y)\,\psi\\
           &=&4\,(M_r\,f_0+b_x\,f_4+b_y\,f_5+b_z\,f_6)\\
\dot g_{11}&=&2\,(4\,M_r\,f_1-4\,M_g\,f_4+b_y\,f_9-b_z\,f_8)\\
\dot g_{12}&=&2\,(4\,M_r\,f_2-4\,M_g\,f_5+b_z\,f_7-b_x\,f_9)\\
\dot g_{13}&=&2\,(4\,M_r\,f_3-4\,M_g\,f_6+b_x\,f_8-b_y\,f_7)\\
\dot g_{14}&=&-4\,\bar\psi\,(M_g\,\y_0+\vec b\,\vec\y)\,\psi\\
           &=&-4\,(M_g\,f_0+b_x\,f_1+b_y\,f_2+b_z\,f_3)\\
\endary

\end{appendix}


\section*{References}

\begin{thebibliography}{9}
  \bibitem{Teng} L.C. Teng: Concerning n-Dimensional Coupled Motions; NAL-Report FN-229 (1971).
  \bibitem{EdwardsTeng} D.A. Edwards and L.C. Teng; (Cont. to PAC '73) IEEE Trans. Nucl. Sci. Vol 20, Issue 3, (1973), 885-888.
  \bibitem{cyc_paper}  C. Baumgarten; Phys. Rev. ST Accel. Beams. 14, 114201 (2011).
  \bibitem{rdm_paper}  C. Baumgarten; Phys. Rev. ST Accel. Beams. 14, 114002 (2011).
  \bibitem{Dragt} Alex J. Dragt; Ann. N.Y. Acad. Sci. 1045: 291-307 (2005).
  \bibitem{DGL} Dae-Gyu Lee; J. Math. Phys. 36 (1995), 524-530.
  \bibitem{BZL} A.O. Barut, J.R. Zeni and A. Laufer; J. Phys. A: Math. Gen. 27 (1994), 6799-6805.
  \bibitem{Hestenes} D. Hestenes: Space-Time Algebra; Gordon and Breach, New
    York, 1966). See also: arXiv:0802.2728v1.
  \bibitem{AJM} A. J. MacFarlane; Commun. math. Phys. 2 (1966), 133-146.
%%%
  \bibitem{LMC} A. Laub and K. Meyer; Celestial Mechanics 9 (1974), 213-238.
  \bibitem{PvL} Chris Paige and Charles Van Loan; Lin. Alg. Appl. 41 (1981) 11-32.
  \bibitem{vanLoan} C.F. Van Loan; Lin. Alg. Appl. 61 (1984) 233-251.
  \bibitem{BMX} P. Benner, V. Mehrmann and H. Xu; J. Comp. Appl. Math. 86 (1997) 17-43.
  \bibitem{BMX2} Peter Benner, Volker Mehrmann and Hongguo Xu; Numer. Math. 78 (1998), 329-358.
  \bibitem{Coleman} R. Coleman; Math. Comp. Sim. 46 (1998) 117-155.
  \bibitem{Dieci} Luca Dieci; Lin. Alg. Appl. 281 (1998) 227-246.   
  \bibitem{SR} D. Sagan and D. Rubin; Phys. Rev. ST Accel. Beams 2, 074001 (1999).
  \bibitem{YQXX} Yan Qing-you and Xong Xi-wen; Appl. Math. and Mech. Vol. 23, No. 11, 2002.
  \bibitem{BKM} Peter Benner, Daniel Kressner, Volker Mehrmann; Future Generation Computer Systems, Vol. 19 Issue 7, (2003) pp. 1243-1252.
  \bibitem{CM} Christian Mehl; SIAM J. Matrix Anal. Appl. 25, No. 4 (2004), 964-985.   
  \bibitem{Luo} Yun Luo;  Phys. Rev. ST Accel. Beams 7, 124001 (2004).
  \bibitem{MK} M. Kleinsteuber; Lin. Alg. Appl. 430 (2009) 155-173.
  \bibitem{MSW} V. Mehrmann, C. Schr\"oder and D.S. Watkins; Lin. Alg. Appl. 431 (2009) 350-368.
  \bibitem{ABK} S. Agoujil, A.H. Bendbib and A. Kanber; Appl. Num. Math. (APNUM-2489, in Press), 2011.
  \bibitem{BFS} P. Benner, H. Fassbender and M. Stoll; Lin. Alg. Appl. 435 (2011) 578-600.
  \bibitem{CS} M. Corless and R. Shorten; Automatica 47 (2011) 431-442.
  \bibitem{FMM} H. Fassbender, D.S. Mackey and N. Mackey; Lin. Alg. Appl. 332-334 (2001) 37-80.
  \bibitem{stat_paper} C. Baumgarten; http://arxiv.org/pdf/1205.3601 (2012).
%%%
  \bibitem{CMC} R.R. Cordeiro, R.V. Martins and A.L.F. Canova; Cel. Mech. Dyn. Astr. 67 (1997), 215-224.
  \bibitem{Talman} R. Talman: Geometric Mechanics; 2nd Ed., Wiley-VCH Weinheim, Germany, 2007. 
  \bibitem{Okubo} Susumu Okubo; Math. Jap. 41 (1995), 59-79: arXiv:hep-th/9408165v1;
  \bibitem{Scharnhorst} K. Scharnhorst; J. of Math. Phys. 40, No. 7 (1999).
  \bibitem{MHO} K.R. Meyer, G.R. Hall and D. Offin: Introduction to Hamiltonian Dynamical Systems and the N-Body Problem; 2nd. Ed.,
                Springer, New York, 2000.
  \bibitem{Parzen} G. Parzen; IEEE Proceedings of PAC 1995.
  \bibitem{HMMG} J.A. Holt, M.A. Martens, L. Michelotti and G. Goderre; Proceedings of the IEEE Part. Acc. Conf. Dallas 1995, FERMILAB-Conf-95/097.
  \bibitem{Wolski} Andrzej Wolski; Phys. Rev. ST Accel. Beams 9, 024001 (2006).
  \bibitem{Arnold} V.I. Arnold: Mathematical Methods of Classical Mechanics; 2nd Ed., Springer, New York 2010.
  \bibitem{Leach} P.G. Leach: On the theory of time-dependent linear canonical transformations as applied to Hamiltonians 
                  of the harmonic oscillator type; J. of Math. Phys. Vol. 18, No. 8 (1977), pp. 1608-1611.
  \bibitem{Hinterberger} Frank Hinterberger, Physik der Teilchenbeschleuniger (in german), 2. Auflage, Springer, Heidelberg 2008.
  \bibitem{Lax} Peter D. Lax; Courant Inst. (N.Y. Univ.), Rep. NYO-1480-87 (1968); 
                also in: Comm. Pure Appl. Math. Vol. 21, No. 5 (1968), pp. 467-490.
  \bibitem{Lax2} W.-H. Steeb and A. Kunick; Chaos in dynamischen Systemen,
    B.I. Wissenschaftsverlag, Mannheim/Wien/Z\"urich (1989), 2nd ed.
  \bibitem{DNR} A. J. Dragt, F. Neri and G. Rangarajan; Phys. Rev. A, Vol. 45, No. 4 (1992), pp. 2572-2584. 
  \bibitem{Jacobi} C. G. J. Jacobi; J. Reine Angew. Math., 30 (1846), pp. 51-95.
\end{thebibliography}
\end{document}